\documentclass[aps,twocolumn,showpacs,tightenlines,nofootinbib,prl]{revtex4-1}
\usepackage{amsmath,amscd,amssymb}
\usepackage[usenames,dvipsnames]{color}
\usepackage{epsfig}
\usepackage{xfrac}
\usepackage{ulem}
\usepackage{latexsym}
\usepackage{mathrsfs}
\usepackage{graphicx}
\usepackage{eso-pic}
\usepackage{bbm}      
\def\beq{\begin{equation}}
\def\eeq{\end{equation}}
\def\be{\begin{eqnarray}}
\def\ee{\end{eqnarray}}
\newcommand{\dslash}{\partial \hskip -0.5em /}

\newcommand{\zr}[1]{\mbox{\hspace*{#1em}}}

\newcommand{\ID}{\mbox{{\sf 1}\zr{-0.16}\rule{0.04em}{1.55ex}\zr{0.1}}}
\begin{document}

\title{The vacuum nature of the QCD condensates}

\author{H. Reinhardt$^{a)}$ and H. Weigel$^{b)}$}

\affiliation{
$^{a)}$Institute for Theoretical Physics, T\"ubingen University
D--72076 T\"ubingen, Germany\\
$^{b)}$Physics Department, Stellenbosch University,
Matieland 7602, South Africa}

\begin{abstract}
Recently it was claimed that QCD condensates were associated with the internal 
dynamics of the hadrons. We challenge this ``in--hadron'' picture of the QCD 
condensates and show that it conflicts well established concepts and experimental 
facts for the example of the quark condensate. We explicitly demonstrate the 
vacuum nature of the quark condensate by a model calculation.
\end{abstract}


\maketitle

\section{Introduction}
\label{sec:1}

At low energies and baryon densities {\it quantum chromo dynamics} (QCD) is in 
the confined phase where chiral symmetry is spontaneously broken while at high 
temperature or density the quarks and gluons form a correlated plasma in which 
chiral symmetry is restored. The various phases of QCD are characterized by order 
(and disorder) parameters. These order (and disorder) parameters are vacuum 
expectation values of certain gauge invariant products of quark and gluon 
operators and are usually referred to as condensates. The most prominent ones 
are the quark condensate $\langle \bar{q} q \rangle$ and the gluon condensate 
$\langle F_{\mu \nu} F^{\mu \nu} \rangle$. These condensates play an important role 
in the operator product expansion where they incorporate the vacuum properties
and they are central to QCD sum rule calculations. 

In in a series of papers \cite{Brodsky:2008be,Brodsky:2010xf} it was recently 
argued that the strong interaction condensates like the quark and gluon condensate 
are not properties of the QCD vacuum but instead associated with the internal 
dynamics of hadrons. To substantiate this interpretation the authors of those 
papers allude to the light front formulation of spontaneous breaking of chiral 
symmetry \cite{Casher:1974xd} and the description of hadrons in terms of 
Dyson--Schwinger and Bethe--Salpeter equations 
\cite{Roberts:1994dr,Alkofer:2000wg,Fischer:2006ub,Roberts:2007jh}. 
This interpretation has far reaching consequences. If the quark and gluon 
condensates are not properties  of the QCD vacuum but of the hadrons
they do not contribute to the cosmological constant while a naive (unrenormalized) 
inclusion of the condensates would result in a cosmological constant, which is 
about a factor of $10^{56}$ larger  than the currently accepted value. 

In this paper we provide strong arguments that invalidate 
the interpretation of the QCD condensates as ``in--hadron'' properties as proposed 
in ref. \cite{Brodsky:2008be,Brodsky:2010xf}. In addition, we will present a 
model calculation that clearly shows that the translationally invariant quark 
condensate is a vacuum property. However, this condensate is strongly distorted 
inside the hadrons.

\section{The quark condensate as property of the QCD vacuum}
\label{sec:2}

Low energy hadron physics is dominated by chiral dynamics. The concept of 
approximate chiral symmetry for the light quark flavors, being spontaneously 
broken in the ground state with the pseudoscalar mesons, like pions and kaons, 
identified as Goldstone bosons, has been overwhelmingly successful. The order 
parameter of chiral symmetry breaking is the quark condensate: It is non--zero 
in the chiral symmetry broken phase and vanishes (for massless quarks) in the 
chirally symmetric phase. For the moment let us assume that the picture of 
``in--hadron'' condensates put forward in 
ref. \cite{Brodsky:2008be,Brodsky:2010xf} were correct, {\it i.e.}
$\langle \bar{q} q \rangle \neq 0$ inside the hadrons and 
$\langle \bar{q} q \rangle = 0$ outside. 
This would imply that chiral symmetry is spontaneously broken only inside the 
hadrons\footnote{We still believe in the general theorems of thermodynamics that 
strict spontaneous symmetry breaking requires an infinite volume but to analyze 
the arguments of ref. \cite{Brodsky:2008be,Brodsky:2010xf} let us, for the time 
being, assume that spontaneous symmetry breaking could take place in a finite 
volume.}, while the exterior of hadrons is in the chirally symmetric phase. 
Consequently the corresponding Goldstone boson, the pion, could exist only inside 
other hadrons. Outside the lightest particle with pion quantum numbers 
would be massive and degenerate with the $\sigma$. This, of course, contradicts 
the experimental observation. The pion is measured outside the hadrons with a 
light mass (for example in a bubble chamber). For an additional counter--argument 
consider nuclear matter as an ensemble of hadrons. If chiral symmetry were 
spontaneously broken only inside the hadrons then there would be no reason why 
this symmetry is restored when the density of baryons is increased. If the 
condensates were indeed hadron properties, their universal character could not be 
sustained, no mechanism would prevent them to depend on the peculiarities of the 
considered hadron. Furthermore we recall recent indications \cite{McLerran:2007qj} 
that at high density a so--called quarkyonic phase of baryonic matter exists 
where quarks and gluons are still confined but chiral symmetry is restored. It is 
thus very unlikely that confinement restricts chiral symmetry breaking to the
interior of hadrons. The discussion in ref. \cite{Brodsky:2010xf} does, unfortunately, 
not illuminate the specific manner in which confinement would trigger such a 
restriction. The mere feature that condensates are extracted from hadron properties
seems irrelevant. This is true for other global quantities and external fields.

The above considerations of empirical facts strongly suggest 
that the interpretation of condensates as hadron properties is invalid. 
And indeed, currently accepted baryon models manifest the opposite picture. The 
quark condensate is non--zero outside the baryon while it vanishes or is, 
at least, substantially reduced inside the hadrons, {\it i.e.} chiral symmetry is 
spontaneously broken outside the hadrons and at least partially restored inside.
To investigate condensates as eventual hadron properties, the model must
allow for a localized description of hadrons.
We will present such a model calculation in the next section. 

The authors of ref. \cite{Brodsky:2008be,Brodsky:2010xf} also refer to the 
superconductor to support their interpretation that the condensates are confined 
to a finite volume. Of course, an actual superconductor has a finite volume. 
However, compared to the characteristic wave length of the electrons in the 
superconductor the latter can, of course, be considered as infinitely extended. 
The superconductor is actually a quite realistic analog to the QCD vacuum. 
In fact, there is lattice evidence \cite{lattice}
that confinement is realized by the dual Meissner effect implying that the QCD 
vacuum represents a dual superconductor. In this analogy the occupied electron 
levels below the Fermi surface correspond to the occupied quark levels of 
the Dirac sea. Usually the electron (cooper-)pair condensate is spread out 
through the whole superconductor and external magnetic fields are expelled 
by the Meissner effect. If, however, an external magnetic field 
is applied to a type II superconductor and the field 
strength exceeds a certain critical value the magnetic field can penetrate 
the superconductor in form of magnetic vortices (flux tubes). At the position 
of the vortices the superconducting phase is destroyed and the U(1)--symmetry 
is restored. The magnetic vortices have the same effect on the superconductor 
as the valence quarks on the QCD vacuum: Their presence causes a local 
restoration (at least partially) of the symmetry that is spontaneously broken 
in the vacuum.

From the above given arguments and the model calculations presented below we 
conclude that the QCD condensates are indeed vacuum but not hadron properties. 

\section{A Model Calculation}
\label{sec:3}

The quark condensate can be obtained from the functional 
\begin{equation}
\langle \overline{q}q \rangle :=
\frac{\int D[\overline{q}]D[q]\,\overline{q}q \,
{\rm exp}(i\mathcal{A})}
{\int D[\overline{q}]D[q]\, {\rm exp}(i\mathcal{A})}\,
\label{eq:def1}
\end{equation}
with $\mathcal{A}=\mathcal{A}[\overline{q},q]$ being the 
underlying action of the quark field $q$. 

\paragraph{Model definition}
Here we will utilize the Nambu--Jona--Lasinio (NJL) model for the quark 
flavor dynamics \cite{Ebert:1985kz} and consider the soliton picture for baryons, 
see refs. \cite{Alkofer:1995mv,Weigel:2008zz} for comprehensive
reviews on these topics. The two--flavor NJL model for the quark
flavor dynamics contains chirally invariant scalar and pseudoscalar 
self--interactions 
\begin{eqnarray}
\mathcal{L}_{\rm NJL}&=&\overline{q}(i\dslash-m_0)q
\cr && 
+2G\sum_{a=0}^{3}
\left[\left(\overline{q}\frac{\tau_a}{2}q\right)^2
+\left(\overline{q}\frac{\tau_a}{2}i\gamma_5 q\right)^2 \right]\,.
\end{eqnarray}
In this model $G$ is the strength of the quark self--interaction and $m_0$
refers to the (small) current quark mass matrix. We assume
two light flavors $\left(\overline{q}=(\overline{u},\overline{d})\right)$
so that $\tau_{1,2,3}$
are the Pauli matrices and $\tau_0=\ID$. 
The model can be bosonized by standard functional methods resulting in an 
effective theory for scalar ($S$) and pseudoscalar ($P$) matrix fields. 
This yields 
\begin{equation}
\int D[\overline{q}]D[q]\, {\rm exp}\left(i\int d^4x 
\mathcal{L}_{\rm NJL}\right)
=\int D[S]D[P]\,{\rm exp}\left(i\mathcal{A}_{\rm B}\right)
\label{eq:pathint}
\end{equation}
with the bosonized action
\begin{eqnarray}
\mathcal{A}_{\rm B}&=&
\frac{-1}{4G}\int d^4x \,{\rm tr}_{\rm F}\left[(S-m_0)^2+P^2\right]
\cr\cr && \hspace{0.5cm} 
-i\,{\rm Tr}_\Lambda {\rm log} \left[i\dslash - (S+i\gamma_5 P)\right]\,.
\label{eq:action1}
\end{eqnarray}
In the above 
expression ${\rm Tr}$ includes the discretized trace over flavor, color 
and Dirac indices as well as spatial integration. Its subscript indicates
regularization of the quarticly divergent functional. The gluon exchange 
of QCD is modeled by the quark self--interaction, that does not depend on 
the color charge of the quarks. Hence the color trace merely produces a 
factor $N_C$.

We evaluate the path integral, eq.~(\ref{eq:pathint}) in the 
saddle point approximation. That is, we consider the meson configuration
$(S_{\rm sp},P_{\rm sp})$ that minimizes the action, $\mathcal{A}_{\rm B}$, 
subject to a prescribed baryon number. In this framework eq.~(\ref{eq:def1}) 
reads
\begin{equation}
\langle \overline{q}q \rangle =
\frac{\int D[\overline{q}]D[q]\,\overline{q}q \,\,
{\rm exp}\left(i\mathcal{A}_{\rm sp}\right)}
{\int D[\overline{q}]D[q]\, {\rm exp}(i\mathcal{A}_{\rm sp})}\,,
\label{eq:qbarqsp}
\end{equation}
with the saddle point action
\begin{eqnarray}
\mathcal{A}_{\rm sp}&=&\int d^4x\, \overline{q}
\left[i\dslash-(S_{\rm sp}+i\gamma_5P_{\rm sp})\right]q\cr\cr
&&-\frac{1}{4G}\int d^4x \,{\rm tr}_{\rm F}
\left[(S_{\rm sp}-m_0)^2+P_{\rm sp}^2\right]
-\mathcal{A}_0\,.
\label{eq:asp}
\end{eqnarray}
We have subtracted a counterterm, $\mathcal{A}_0$, such that the saddle point 
action (density) vanishes for the vacuum configuration of the meson fields. 
In the language of Feynman diagrams $\mathcal{A}_0$ is the fermion loop 
without external legs. As in any flat--space field theory it is purely kinematical 
and dispensed of any dynamics. Similarly to the local integral it exactly 
cancels\footnote{In the current saddle point approximation this cancellation 
occurs inevitably in contrast to its innuendo as a mathematical mistake in 
ref. \cite{Brodsky:2012ku}.} in the ratio, eq.~(\ref{eq:qbarqsp}). Standard 
functional techniques yield
\begin{equation}
\langle \overline{q}q \rangle =
\frac{\delta}{\delta J} \Big\{i\,{\rm Tr}_\Lambda {\rm log}
\left[i\dslash-J-(S_{\rm sp}+i\gamma_5P_{\rm sp})\right]\Big\}_{J=0}\,.
\label{eq:source}
\end{equation}
In what follows we will omit the subscripts for the saddle point
approximation.

Since the scalar fields couple to $\overline{q}\tau_a q$, we do not have to 
take the detour, eq.~(\ref{eq:source}), of introducing an external source 
but may directly compute
\begin{equation}
\langle \overline{q}_iq_j \rangle = \frac{\delta}{\delta S_{ij}}
\Big\{i\,{\rm Tr}_\Lambda {\rm log}
\left[i\dslash - (S+i\gamma_5 P)\right]\Big\}\,.
\label{eq:def2}
\end{equation}

\paragraph{Vacuum and meson sector}
\label{sec:vacuum}

The ground state of the model is obtained from the variational 
principle $\frac{\delta \mathcal{A}_{\rm B}}{\delta S,P}=0$.  Symmetry arguments 
require the ground state to be flavor symmetric and homogenous, 
$\langle S \rangle=m \ID$, and $\langle P \rangle=0$. Since any 
non--zero vacuum expectation value of the scalar field provides mass for 
the fermions, $m$ is called the constituent quark mass. The variation 
with respect to the scalar field $S_{11}$ results in the gap--equation 
$m=m_0-2G\langle\bar{u}u\rangle_{\rm V}$ that embodies the vacuum
up--quark condensate which upon {\it proper time} 
regularization~\cite{Ebert:1985kz} becomes
\begin{equation}
\langle \overline{u}u\rangle_{\rm V}=
-\frac{N_C}{4\pi^2}\,m^3\,\Gamma\left(-1,\frac{m^2}{\Lambda^2}\right)\,.
\label{eq:meson}
\end{equation}
Of course, isospin symmetry yields 
$\langle\overline{u}u\rangle_{\rm V}=\langle\overline{d}d\rangle_{\rm V}$.
Pion properties are determined by expanding the action to quadratic order 
in $P$. The pole condition for the resulting propagator relates the model 
parameters to the pion decay constant and the pion mass \cite{Ebert:1985kz}
\begin{equation}
f^2_\pi=m^2 \frac{N_C}{4\pi^2}\int_0^1dx\
\Gamma\left(0,\frac{m^2-x(1-x)m_\pi^2}{\Lambda^2}\right)
\label{eq:fpi}
\end{equation}
and
\begin{equation}
m_\pi^2f_\pi^2=\frac{m_0 m}{G}\,.
\label{eq:mpi}
\end{equation}
Then the gap--equation can be framed as 
\begin{equation}
m=m_0\left[1+\frac{N_C\, m^4}{2\pi^2 f_\pi^2m_\pi^2}\,
\Gamma\left(-1,\frac{m^2}{\Lambda^2}\right)\right]\,.
\label{eq:gap2}
\end{equation}
Given a value for the constituent quark mass $m$, the 
equations~(\ref{eq:fpi}), (\ref{eq:mpi})
and~(\ref{eq:gap2}) determine the model parameters from $f_\pi=93{\rm MeV}$ and 
$m_\pi=135{\rm MeV}$. In turn, the quark condensate is computed from 
equation~(\ref{eq:meson}). The above equations refer to the physical case 
of non--zero pion mass. The chiral limit $m_0 \to 0$ can be smoothly approached. 
Then eq.~(\ref{eq:mpi}) is trivially fulfilled and eq.~(\ref{eq:gap2})
turns into $m=(N_cm^3/2\pi^2)\,G\,\Gamma(-1,m^2/\Lambda^2)$, 
fixing the coupling $G$ for given values of $m$ 
and $\Lambda$, {\it cf.} eq.~(\ref{eq:fpi}) .

In this treatment pions are considered as plane waves. Constructing
localized objects as a wave--packet does not alter any of the 
above equations. In particular, the condensate, eq.~(\ref{eq:meson}) 
remains translationally invariant.

\paragraph{Baryon as soliton}
\label{sec:soliton}

The soliton picture for baryons is well established and 
successful~\cite{Weigel:2008zz}. For the current objective it is 
particularly attractive because the large $N_C$ expansion\footnote{Here we
omit subleading contributions. They arise from fluctuations about the soliton.}
establishes a framework to consider baryons as localized objects that avoids 
the introduction of wave--packets. The solitons \cite{Reinhardt:1988fz} of the
NJL--model are discussed at length in ref.~\cite{Alkofer:1994ph}. They are 
static configurations of the matrix fields $S$ and $P$ and give rise to a 
stationary Dirac Hamiltonian, $h$ with single particle energy eigenvalues 
$\epsilon_\nu$. From these the static energy functional is obtained 
as~\cite{Reinhardt:1989st},
\begin{equation}
E_F=N_C\theta(\epsilon_{\rm val})\, \epsilon_{\rm val}
+\frac{N_C}{4\sqrt{\pi}}\sum_\nu |\epsilon_\nu| \Gamma\left(-\frac{1}{2},
\left(\frac{\epsilon_\nu}{\Lambda}\right)^2\right)\,,
\label{eq:eng}
\end{equation}
again, for {\it proper time} regularization. The valence level (val) is the 
state of largest binding due to the static background (soliton) field. If its energy 
is positive, explicit occupation provides baryon number. Otherwise, the 
polarized vacuum carries the baryon charge.  
The valence quark contribution is fully contained in $\mathcal{A}_B$ since
the prescribed baryon number determines the sum of the occupation numbers that
occur in the computation of the functional integral, eq.~(\ref{eq:pathint})
\cite{Alkofer:1994ph,Reinhardt:1989st}. Minimizing 
the total energy singles out the level of largest binding.

For simplicity, and to avoid the subtle issue of scalar instabilities in this
model, we consider the hedgehog configuration on the chiral circle. This
results in the Dirac Hamiltonian
\begin{equation}
h=\vec{\alpha}\cdot\vec{p}+m\beta \big[
{\rm cos}\Theta(r)+i\gamma_5\vec{\tau}\cdot\hat{r}\,
{\rm sin}\Theta(r)\big]\,.
\label{eq:Dirac}
\end{equation}
The chiral angle $\Theta(r)$ self--consistently minimizes the total 
energy functional 
\begin{equation}
E[\Theta]=E_F+4\pi m_\pi^2f_\pi^2\int_0^\infty dr r^2\,
\left[1-{\rm cos}(\Theta(r))\right] \, .
\end{equation}
The soliton can also be constructed in the chiral limit.
No qualitative differences emerge~\cite{Alkofer:1994ph}.

Similar to the vacuum sector we find the quark condensate from a 
functional derivative of the action (energy) in the baryon number 
one sector. Assuming formally that
${\rm cos}\Theta$ and ${\rm sin}\Theta$ are independent fields, the
condensate in the soliton (baryon) background is obtained from 
$\frac{\delta E_F}{\delta {\rm cos}\Theta}$ to be
\begin{eqnarray}
\langle \overline{u}u\rangle_{\rm S}&=&\frac{N_C}{2}\int \frac{d\Omega}{4\pi}
\Big\{\theta(\epsilon_{\rm val})
\Psi_{\rm val}^\dagger(\vec{r\,})\beta \Psi_{\rm val}(\vec{r\,})
\cr \cr && 
-\frac{1}{2}\sum_\nu {\rm sign}(\epsilon_\nu)\,
{\rm erfc}\left(\frac{\epsilon_\nu}{\Lambda}\right)
\Psi_\nu^\dagger(\vec{r\,})\beta \Psi_\nu(\vec{r\,})\Big\}\,.
\qquad
\label{eq:totscal}
\end{eqnarray}
The additional factor $\frac{1}{2}$ arises because the functional trace from 
which $E_{\rm F}$, eq.~(\ref{eq:eng}) is extracted includes isospin and
$\langle \overline{u}u\rangle_{\rm S}=\langle \overline{d}d\rangle_{\rm S}$ 
in leading order of the $1/N_C$ expansion. 
The computation of $\langle \overline{u}u\rangle_{\rm S}$ in this model is 
not fully new ({\it cf.} refs. in \cite{Alkofer:1994ph} and \cite{Wakamatsu:1992wr}). 
Here we particularly focus on its relation to the vacuum sector.

The relative simplicity of the expressions for the quark condensates,
eqs.~(\ref{eq:meson}) and~(\ref{eq:totscal}) make this model a perfect
candidate to examine recent ideas on condensates as hadron 
properties~\cite{Brodsky:2008be,Brodsky:2010xf}. 

\paragraph{Numerical Results}
\label{sec:numerics}

In the numerical evaluation of the above expressions we consider the 
physical value $N_C=3$, though our analysis refers to the large $N_C$ limit.

\begin{figure}[t]
\centerline{
\epsfig{file=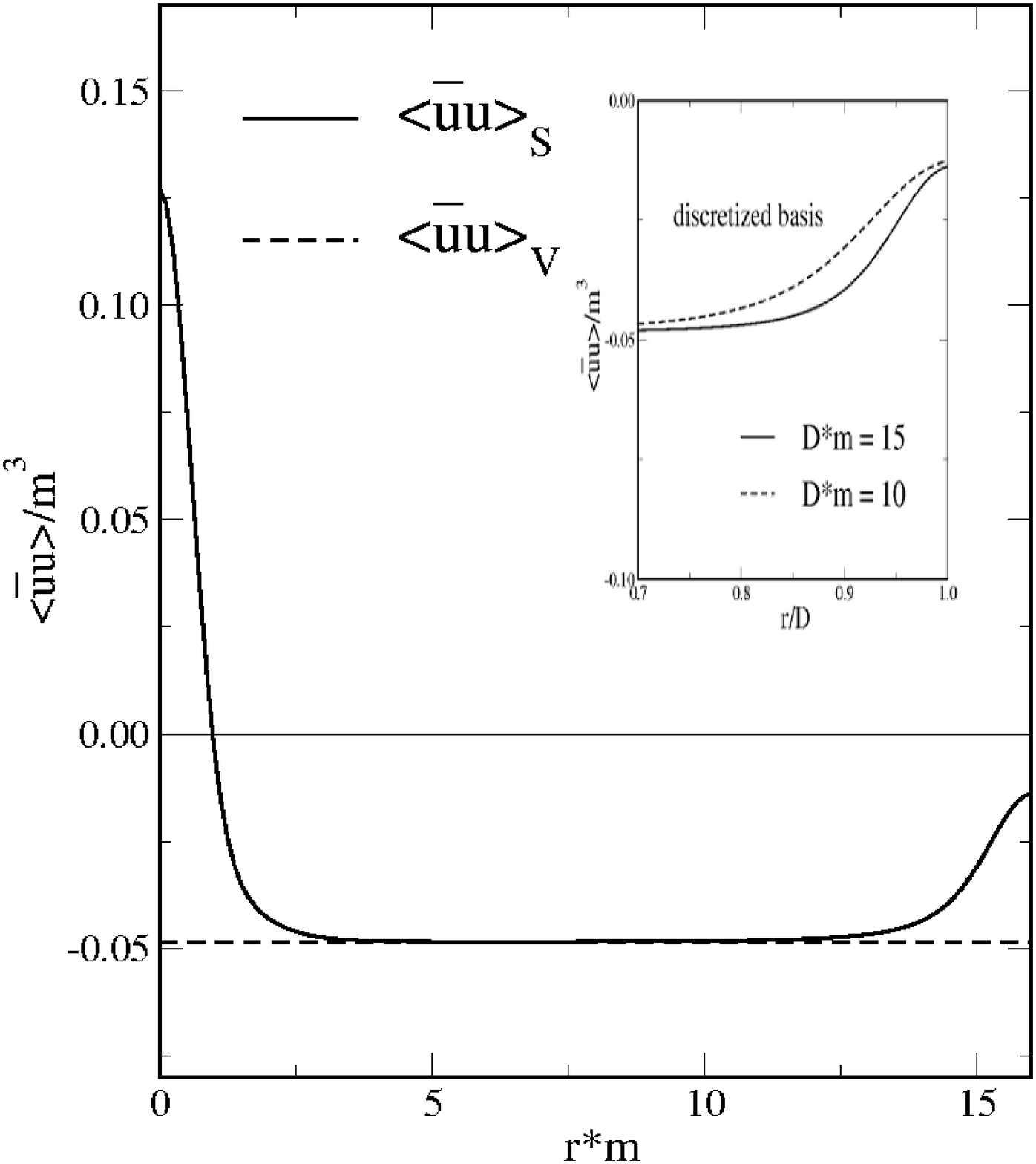,width=8.0cm,height=6.0cm}}
\caption{\label{fig_tot}\sf Total scalar condensate, eq.~(\ref{eq:totscal})
as a function of the distance from the center of the soliton. The dashed
line indicates the vacuum/meson values for the condensate, eq.~(\ref{eq:meson}).
Dimensions are measured in appropriate powers of the constituent quark mass,
$m=450{\rm MeV}$. The inserts illuminate finite size effects by varying 
the scale $D$ at which the eigenstates are discretized.}
\end{figure}

We display numerical results for a typical value of $m=450{\rm MeV}$ in 
figure~\ref{fig_tot}. This value of the constituent quark mass is large 
enough to provide dynamical binding\footnote{This subtraction is reminiscent of 
$\mathcal{A}_0$ in eq.~(\ref{eq:asp}).}, {\it i.e.} $E[\Theta]-E[0]<N_C m$.
The soliton is localized in the regime $r\le 2/m$. Outside that regime
the condensate equals that of the meson sector, {\it i.e.}, the non--zero
translationally invariant vacuum result. We observe finite size effects as 
mitigations of the condensate far away from the soliton. Obviously the 
region of deviation from the non--zero vacuum value decreases as the 
(unphysical) system size is increased. Definitely, this analysis shows 
that the mitigation at large distances is a mere artifact of the numerical 
procedure that makes no statement about the actual behavior of the condensate.

\begin{figure}[t]
\centerline{~~~~
\epsfig{file=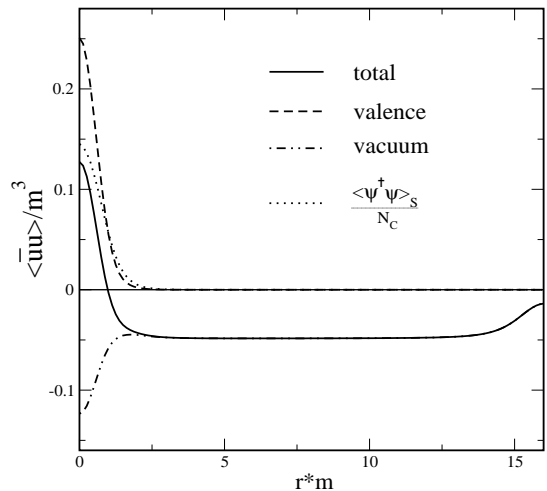,width=8.0cm,height=8.0cm}}
\caption{\label{fig_sep}\sf 
Separation of quark condensate in valence and vacuum
contributions according to eq.~(\ref{eq:totscal}). 
The dotted line is the baryon charge density.}
\end{figure}

As seen from figure~\ref{fig_sep} inside the soliton regime the condensate 
is dominated by the valence quark contribution. This contribution actually 
over--compensates the vacuum result, which indeed tends to increase (in magnitude) 
in region of the baryon. This is a clear indication that it is actually the 
in--hadron region, where the condensates are modified by the interactions that 
bind quark to hadrons. On the contrary the hadronless regime possesses constant 
non--zero condensates as the comparison with baryon charge density clearly 
demonstrates. The condensate persists in areas of zero charge density but 
is distorted in the domain of the hadron. This reverses the conjecture 
of refs.~\cite{Brodsky:2008be,Brodsky:2010xf}. 


In the sprit of refs.~\cite{Brodsky:2008be,Brodsky:2010xf} one might
challenge this model for its lack of confinement. As far as the existence of
asymptotic quark states concerns, confinement is indeed lacking. Note however,
that matrix elements of any non--singlet color operator vanish (by averaging)
in the model. In addition to the above observed non--correlation between baryon
charge density and condensate this suggests that confinement is not essential for
the existence of condensates. It is generally accepted that confinement is
due to the gluon sector of QCD, while chiral symmetry breaking is due to
the interaction between the quarks (mediated by the gluons).

~\smallskip
\section{Conclusion}
\label{sec:4}

We have explained that recent interpretations of QCD condensates being
''in--hadron'' quantities contradict the phenomenology of low--energy
strong interactions. Furthermore we have presented a model calculation
for the quark condensate that strongly supports the understanding that
these condensates are vacuum properties that are significantly distorted
or reduced in hadrons. We recall that this picture is fully affirmed by 
lattice QCD measurements of the quark condensate in the presence of static 
sources~\cite{Buerger:1993bq}.

Refs~\cite{Brodsky:2008be,Brodsky:2010xf} suggest that the in--hadron 
nature of condensates were due to QCD being a confining theory, a property not 
manifest in our model calculation. Yet the explicit formulas in those 
papers do not reveal the specific role of confinement. To a large extend
ref.~\cite{Brodsky:2010xf} relies on relations between matrix elements of
QCD operators, the QCD vacuum and a momentum eigenstate with pion quantum
numbers. These relations result from chiral symmetry and are also valid
in our model. In ref~\cite{Brodsky:2008be} the finite range of the hadron
wave--function in the light--front formulation is crucial. While there is 
no straightforward relation between light--front and instantaneous
wave--functions~\cite{Miller:2009fc}, the current model calculation does 
not indicate that the range of the quark condensate is related to that
of the nucleon wave--function (represented by the baryon density).

Apparently the subtle puzzle of the too large contribution from 
condensates to the cosmological constant remains. We note, however,
that the non--zero condensates are quantum effects, as is the zero point
energy contribution to the cosmological constant. This suggests that any 
solution to the puzzle ultimately requires a consistent formulation of quantum 
gravity rather than reinterpreting entries of the classical energy momentum tensor.

\subsection{Acknowledgments}
We acknowledge comments from the authors of ref.~\cite{Brodsky:2010xf} 
on a draft version of this manuscript. One of us (HW) acknowledges 
correspondence with R.L. Jaffe. We thank T. Sch\"afer for substantiating 
and encouraging comments on the first version of this manuscript. This 
work is supported in parts by the German Science Foundation (DFG-Re 856/6-3)
and the National Research Foundation of South Africa.

\end{document}